\documentclass[vecphys]{svmult}

\usepackage{graphicx}        
\usepackage{times,amsmath}
\usepackage{color}

\usepackage{multicol}        

\makeindex             

\newcommand{\fc}{f_{\mbox{\scriptsize  c}}}
\newcommand{\kT}{k_{\mbox{\scriptsize B}}T}

\newcommand{\half}{\frac{1}{2}}
\newcommand{\Ffriction}{F_{\mbox{\scriptsize  friction}}}
\newcommand{\Fthermal}{F_{\mbox{\scriptsize  thermal}}}
\newcommand{\tildeFtherm}{\tilde{F}_{\mbox{\scriptsize thermal}}}
\newcommand{\tildeFfric}{\tilde{F}_{\mbox{\scriptsize friction}}}
\newcommand{\Fext}{F_{\mbox{\scriptsize  external}}}
\renewcommand{\Re}{\mbox{Re}}

\newcommand{\fnu}{f_{\nu}}

\newcommand{\gstokes}{\gamma_{\mbox{\scriptsize Stokes}}}

\newcommand{\vecv}{\vec{v}}
\newcommand{\vecx}{\vec{x}}
\newcommand{\vecd}{\vec{d}}
\newcommand{\veceta}{\vec{\eta}}
\newcommand{\vecnull}{\vec{0}}
\newcommand{\vstage}{v_{\mbox{\scriptsize stage}}}
\newcommand{\fstage}{f_{\mbox{\scriptsize stage}}}
\newcommand{\fsample}{f_{\mbox{\scriptsize sample}}}
\newcommand{\ndim}{n_{\mbox{\scriptsize dim}}}

\catcode`Ï=\active \defÏ{{\Large$\bullet$}}
\catcode`å=\active \defå{{\aa}}
\catcode`Å=\active \defÅ{{\AA}}
\catcode`æ=\active \defæ{{\ae}}
\catcode`Æ=\active \defÆ{{\AE}}
\catcode`ø=\active \defø{{\o}}
\catcode`Ø=\active \defØ{{\O}}
\catcode`ä=\active \defä{{\"a}}
\catcode`ö=\active \defö{{\"o}}
\catcode`ü=\active \defü{{\"u}}
\catcode`ó=\active \defó{{\'{o}}}
\catcode`é=\active \defé{{\'{e}}}
\catcode`à=\active \defà{{\`{a}}}
\catcode`á=\active \defá{{\'{a}}}
\begin{document}

\title*{Brownian Motion after Einstein:\\ Some new applications and new experiments}
\titlerunning{Brownian Motion after Einstein} 
\author{D. Selmeczi\inst{1,2}\and
S. F. Toli\'c-Nørrelykke\inst{3}\and
E. Schäffer\inst{4}\and
P. H. Hagedorn\inst{5}\and
S. Mosler\inst{1}\and
K. Berg-Sørensen\inst{6,7}
\and
N. B. Larsen\inst{1,5}\and
H. Flyvbjerg\inst{1,5,8}
}
\authorrunning{D. Selmeczi et al.} 
\institute{
$^1$\ Danish Polymer Centre, Ris{\o} National Laboratory, DK-4000 Roskilde, Denmark
\\
$^2$\ Department of Biological Physics, Eötvös Loránd University (ELTE),
                                                          H-1117 Budapest, Hungary
                                                          \\
$^3$\  Max Planck Institute for the Physics of Complex Systems,
 Nöthnitzer Strasse 38, D-01187, Dresden, Germany
\\
$^4$\  Max Planck Institute for Molecular Cell Biology and Genetics,
 Pfotenhauer Strasse 108, D-01307 Dresden, Germany
 \\
$^5$\ Biosystems Department, Ris{\o} National Laboratory, DK-4000 Roskilde, Denmark
\\
$^6$\  The Niels Bohr Institute, 
Blegdamsvej 17, DK-2100 Copenhagen Ø, Denmark
\\
$^7$\ Current address: 
Department of Physics, Technical University of Denmark, Building 309, DK-2800 Kgs.~Lyngby, Denmark
\\ 
$^8$\  Corresponding author, henrik.flyvbjerg@risoe.dk
}

\maketitle

\section{Introduction}

The first half of this chapter describes the
development in mathematical models of Brownian motion 
after Einstein's seminal papers \cite{einstein56}
and  current applications to optical tweezers.  
This instrument of choice among single-molecule biophysicists 
is also an instrument of precision that requires an understanding 
of Brownian motion beyond Einstein's.
This is illustrated with some applications, current and potential, 
and it is  shown how addition of a controlled forced motion 
on the nano-scale of the tweezed object's  thermal motion 
can improve the calibration of the instrument in general,
and make it possible also in complex surroundings.
The second half of the present chapter, starting with Sect.~9,  
describes the co-evolution of biological motility models 
with models of Brownian motion, including very recent results for how to derive
cell-type-specific motility models from experimental cell trajectories.

\section{Einstein's Theory}
\label{intro}
When Einstein in 1905 formulated the theory that quickly became known as his theory for
Brownian motion, he did not know much about this motion\footnote{Just how much he 
 knew seems an open question that may never be answered \cite{Renn2005}.}
He was looking for observable consequences of what was then called
\textit{the molecular-kinetic theory of heat}.
So he was not concerned about the finer details of specific situations.
In fact, apart from dated mathematical language,
his papers on Brownian motion \cite{einstein56} remain paradigms for
how to model the essence of a phenomenon with ease and
transparency by leaving out everything that can possibly be left out.

The simplest version of his theory, 
\begin{equation}
 \dot{x}(t) = (2D)^{\half}  \, \eta(t) \enspace,
\label{eq:einstein}
\end{equation}
for the trajectory $x(t)$ of a Brownian particle,
here in one dimension and in the language of Langevin \cite{Langevin_1908,Lemons_Gythiel_1997},
works so well also for real experimental situations
that its extreme simplicity may be overlooked:
No simplification of this theory is possible.
The white noise $\eta(t)$ is the simplest possible:
\begin{equation}
\mbox{For all } t,\,t', ~~~ \langle \eta(t) \rangle = 0  \mbox{~~and~~}
\langle \eta(t) \eta(t')\rangle = \delta(t-t') \enspace.
\label{eq:whitenoise}
\end{equation}
When this noise is normalized as done here---as simple as possible---%
the dimensions of $x$ and $\eta$ require 
that a constant with  dimension of diffusion coefficient appears where it does in (\ref{eq:einstein}).
Equation~(\ref{eq:einstein}) is mathematically equivalent to the diffusion equation, 
introduced by Fick in 1857, in which the diffusion coefficient $D$ is already defined,
and that determines the factor $2D$ in (\ref{eq:einstein}).
The new physics was in Einstein's assumption that Brownian particles also diffuse,
and in his famous relation, the fluctuation-dissipation theorem
\begin{equation}  \label{eq:einsteinsrelation}
D = \kT/\gamma_0 \enspace,
\end{equation}
which relates their diffusion coefficient $D$ and their Stokes' friction coefficient $\gamma_0$
via the Boltzmann energy $\kT$.
It is derived by introducing a constant external force field in (\ref{eq:einstein}),
and assuming Boltzmann statistics in equilibrium.
For a spherical particle,
\begin{equation} \label{eq:stokeslaw}
\gamma_0 = 6\pi \rho \nu R
\enspace,
\end{equation}
where  $\rho$ is the density of the fluid,
$\nu$ its kinematic viscosity,
and $R$ is the sphere's radius.

\section{The Einstein-Ornstein-Uhlenbeck Theory}

Details left out in the model described in (\ref{eq:einstein}--\ref{eq:stokeslaw})
will be found missing, of course, if one looks in the right places.
For example, the length of the trajectory $x(t)$
is infinite for any finite time interval considered\footnote{Consider an interval of duration $t$.
Split it into $N$ intervals of duration $\Delta t=t/N$.
In each of these, the mean squared displacement of the Brownian particle is $2D\Delta t$.
So on the average, the distance travelled in a time interval 
of duration $\Delta t$ is proportional to $(\Delta t)^{1/2}\propto N^{-1/2}$.
Consequently, the distance travelled in a time interval 
of duration $t$ is proportional to $t^{1/2}\propto N^{1/2}$.
Let $N\rightarrow\infty$, and the infinite trajectory has been demonstrated.
The proof can be made mathematically rigorous in the formalism of Wiener processes, 
e.g., which is just the mathematical theory of Brownian motion.}.
Ornstein and Uhlenbeck \cite{ornstein18,uhlenbeck30} showed that this mathematical absurdity
does not appear in  Langevin's equation \cite{Langevin_1908},
\begin{equation} \label{eq:Langevin}
m\ddot{x}(t) = - \gamma_0 \dot{x}(t) + \Fthermal(t)
\enspace,
\end{equation}
where $m$ is the inertial mass of the Brownian particle,
and the force from the surrounding medium
is written as a sum of two terms:
Stokes friction, $- \gamma_0 \, \dot{x}$,
and a random thermal force $\Fthermal = (2\kT \gamma_0)^{1/2} \, \eta(t)$
with ``white noise'' statistical properties following from (\ref{eq:whitenoise}).
The random motion resulting from (\ref{eq:Langevin})
is known as the {\em Ornstein-Uhlenbeck process} (OU-process).
In the limit of vanishing $m$, Einstein's theory is recovered.
Together, they make up the Einstein-Ornstein-Uhlenbeck theory of Brownian motion.

The OU-process improves Einstein's simple model
for Brownian motion by taking the diffusing particle's inertial mass into account.
As pointed out by Lorentz \cite{Lorentz_21}, however,
this theory is physically correct 
only when the particle's density is much larger than the fluid's.
When particle and fluid densities are comparable,
as in the motion Brown observed,
neither Einstein's theory nor the OU-process are consistent with hydrodynamics.
This is seen from exact results by Stokes from 1851 
and by Boussinesq from 1903 
for the force on a sphere that moves with \textit{non}-constant velocity,
but vanishing Reynolds number, through an incompressible fluid. 
Hydrodynamical effects that the OU-process ignores,
are more important than the inertial effect of the particle's mass.
These effects are the frequency-dependence of friction
and the inertia of entrained fluid.
Stokes obtained the friction coefficient, (\ref{eq:stokeslaw}),
for motion with constant velocity \cite{Stokes1851}.
Brownian motion is anything but that.
Also, mass and momentum of the fluid entrained
by a sphere doing rectilinear motion with constant velocity is \textit{infinite}
according to Stokes solution to Navier-Stokes equation \cite{Stokes1851,LandauLifshitz1959}.
This gives a clue that entrained fluid matters,
and the pattern of motion too.

But since Einstein's theory explained experiments well,
this hydrodynamical aspect of Brownian motion did not demand attention.
Not until computers made it possible to simulate molecular dynamics.


\section{Computer Simulations: More Realistic than Reality}

In 1964--66 Rahman simulated liquid Argon as a system of spheres 
that interacted with each other through a  Lennard-Jones potential~\cite{Rahman_1964,Rahman_1966}.
He measured a number of properties of this simple liquid, including the 
velocity auto-correlation function $\phi(t) = \langle \vecv(t)\cdot\vecv(0)\rangle $,
which showed an initial rapid decrease, followed by a slow approach to zero from below, 
i.e., there was a negative long-time tail.
Several attempts were made to explain his results theoretically, with mixed success.

In the years 1967--1970 Alder and Wainwright  simulated liquid Argon 
as a system of hard spheres and observed hydrodynamic patterns in the movement of 
spheres surrounding a given sphere, 
though all the spheres supposedly did Brownian motion~\cite{Alder_Wainwright_67,Alder1970}.
Using a simple hydrodynamical dimension argument, 
and supporting its validity with numerical solutions to Navier-Stokes equations,
they argued that the velocity auto-correlation function has a positive power-law tail,
 $\phi(t)\propto t^{-3/2}$ in three-dimensional space.
This result is in conflict with the velocity auto-correlation function for the OU-process, 
which decreases exponentially, with characteristic time $m/\gamma$.
But the 3/2 power-law tail agrees also with Alder and Wainright's  
simulation results for a simple liquid of hard spheres
doing Brownian motion.

This made theorists \cite{zwanzig_bixon_70} remember Stokes' result from 1851 
for the friction on a sphere that moves with \textit{non}-constant velocity:
There are actually \textit{two} Stokes' laws, published in the same paper~\cite{Stokes1851}.
Einstein had used the simplest one, the one for movement with constant velocity,
so the effect of accelerated motion is not accounted for in his theory.
Nor is it in the Ornstein-Uhlenbeck theory.
However, acceleration of a particle in a fluid also accelerates the fluid surrounding the particle,
in a vortex ring (in three dimensions, and two vortices in two dimensions) that persists for long, 
disappearing only by broadening at a rate given by the kinematic viscosity~\cite{Alder1970}.
In this way the fluid ``remembers" past accelerations of the particle.
This memory affects the friction on the particle at any given time
in a manner that makes the dynamics of the particle depend on its past
more than inertial mass can express.
The result is an effective dynamical equation for the particle, Newton's Second Law with
a memory kernel, as we shall see.


\section{Stokes Friction for a Sphere in \textit{Harmonic} Rectilinear Motion}
\label{sec:hydrodynamicmodel}

The friction coefficient
that is relevant for a more correct description of Brownian motion,
differs from the friction coefficient
that most often is associated with Stokes' name, (\ref{eq:stokeslaw}),
but it is actually the main subject of reference~\cite{Stokes1851}.
Stokes was not addressing the hydrodynamics
of Brownian motion in 1851,
but the hydrodynamics of an incompressible fluid
surrounding a sphere
that does rectilinear \textit{harmonic} motion
with no-slip boundary condition,
at vanishing Reynolds number,
and with the fluid at rest at infinity.
The equations describing this motion are linear, however,
and \textit{any} trajectory of a particle
can be written as a linear superposition of harmonic trajectories,
by virtue of Fourier analysis \cite{Boussinesq_1903}.
So the flow pattern around a sphere following any trajectory
can be written as a superposition of flows around spheres in harmonic motion,
as long as the condition of vanishing Reynolds number
is satisfied by the arbitrary trajectory.
It is for a Brownian particle's trajectory,
so Stokes' result for harmonic motion
is fundamental for the correct description of Brownian motion.

In general, the instantaneous friction experienced 
by a rigid body that moves through a dense fluid like water,
depends on the body's past motion,
since the past motion determines the fluid's present motion.
For a sphere performing rectilinear harmonic motion $x(t;f)$
with cyclic frequency $\omega=2\pi f$ in an incompressible fluid
and at vanishing Reynolds number,
Stokes found the ``frictional'' force 
\cite{Stokes1851},\cite[\S 24, Problem 5]{LandauLifshitz1959},
\begin{eqnarray}
\Ffriction(t;f) &=& - \gamma_0 \left( 1 + \frac{R}{\delta(f)}\right) \dot{x}(t;f)  \nonumber \\
 && -\left(3\pi \rho R^2 \delta(f) + \frac{2}{3}\pi \rho R^3 \right) \ddot{x}(t;f)   \label{eq:Ffricmod} \\
&& = -\gstokes(f) \, \dot{x}(t;f) \enspace;  \nonumber\\
\gstokes(f) &\equiv& \gamma_0 \left( 1 + (1-i) \frac{R}{\delta(f)} - i \,\frac{2R^2}{9\delta(f)^2}\right)
\enspace,
\end{eqnarray}
where only the term containing $\dot{x}(t;f)=-i2\pi f x(t;f)$ dissipates energy,
while the term containing $\ddot{x}(t;f) = -(2\pi f)^2 x(t;f)$ 
is an inertial force from entrained fluid.
The notation is the same as above:
$\gamma_0$ is the friction coefficient
of Stokes' law for rectilinear motion with constant velocity,
(\ref{eq:stokeslaw}).
The {\em penetration depth\/} $\delta$ characterizes the exponential decrease
of the fluid's velocity field as function of distance from the
oscillating sphere.
It is frequency dependent,
\begin{equation} \label{eq:delta}
\delta(f) \equiv (\nu/\pi f)^{\half} =   R (\fnu/f)^{\half}
\enspace,
\end{equation}
and large compared to $R$ for the frequencies we shall consider.
For a sphere with diameter $2R=1.0\,\mu$m in water at room temperature
where $\nu=1.0\,\mu$m$^2/\mu$s,
$\fnu \equiv \nu/(\pi R^2) = 1.3$\,MHz.

Note that the mass of the entrained fluid, the coefficient to $\ddot{x}$ in (\ref{eq:Ffricmod}),
becomes infinite in the limit of vanishing frequency $f$,
i.e., the flow pattern around a sphere moving with constant velocity has infinite momentum, 
according to Stokes' steady-state solution to Navier-Stokes' equations.


\section{Beyond Einstein: Brownian Motion in a Fluid}
\label{sect:brownianincompressible}

The friction on a sphere that, without rotating,
follows an arbitrary trajectory $x(t)$ with vanishing Reynolds
number in an incompressible fluid that is at rest at infinity,
is found by Fourier decomposition of $x(t)$ to a superposition
of rectilinear oscillatory motions $\tilde{x}(f)$.
Using (\ref{eq:Ffricmod}) on these, gives
\begin{equation}
\tildeFfric(f) = -\gstokes(f) (-i2\pi f) \, \tilde{x}(f) \enspace,
\end{equation}
which Fourier transforms back to~\cite{Boussinesq_1903},
\begin{eqnarray} \label{eq:retardedfriction}
\lefteqn{\Ffriction(t) = -\gamma_0 \, \dot{x}}\\
&&-\frac{2}{3}\pi \rho R^3 \, \ddot{x}(t)
-6\pi\rho R^3 \fnu^{1/2} \int_{-\infty}^t dt' (t-t')^{-1/2} \, \ddot{x}(t')
\enspace.\nonumber
\end{eqnarray}
So the Langevin equation (\ref{eq:Langevin}) is replaced by \cite{widom71,case71}
\begin{equation} \label{eq:hydro}
m\ddot{x}(t) = \Ffriction(t) + \Fext(t) + \Fthermal(t)  \enspace,
\end{equation}
where $\Fext$ denotes all external forces on the sphere,
such as gravity or optical tweezers,
and $\Fthermal$ denotes the random thermal force on the sphere from
the surrounding fluid.

Several authors have derived expressions for the thermal force
using different arguments and finding the same result
\begin{equation}  \label{eq:hydronoise}
\tildeFtherm(f)= \left( 2\kT\, \Re\gstokes(f) \right)^{\half} \tilde{\eta}(f)
\enspace;
\end{equation}
see overviews in \cite{Bedeaux1974,Pomeau_Resibois_1975}
\footnote{Here we have written the frequency-dependent noise amplitude explicitly, 
and to this end introduced $\tilde{\eta}(f)$, 
the Fourier transform of a white noise $\eta(t)$,
normalized as in (2).}.
Briefly, Brownian motion in a fluid
is the result of fluctuations in the fluid described
by fluctuating hydrodynamics~\cite[Chapter XVII]{LandauLifshitz1959}\footnote{Readers familiar 
with the Green-Kubo theory of linear response to perturbations 
may appreciate fluctuating hydrodynamics as a case where the order 
of \emph{linearization} and ``\emph{stochastization}" \cite[Sect.~4.6]{Kubo_Toda_Hashitsume_1978} 
is a non-issue by virtue of the Reynolds number for thermal fluctuations.}.
In this theory one assumes that the random currents split up
into systematic and random parts,
the former obeying (Navier-)Stokes equation, the latter obeying a fluctuation-dissipation theorem.
From this theory one derives the expression of the thermal force
on a sphere in the fluid.

Note that this description did \emph{not} invoke a scenario of randomly moving molecules 
that bump into the micro-sphere and thus cause its Brownian motion.
This scenario is correct for Brownian motion \emph{in a dilute gas}. 
It is of great pedagogical value in undergraduate teaching.
But it does not apply to fluids! 
The scientific literature shows that some undergraduates proceed to become scientists 
without realizing this limitation on the scenario's validity.
However, the coarse-grained description that replaces a molecular description with a hydrodynamical one,
is a very good approximation on the length- and time-scales of the 
thermal fluctuations that drive the Brownian motion of a micron-sized sphere \emph{in a fluid}.  
This is why fluctuating hydrodynamics \cite[Chapter XVII]{LandauLifshitz1959} 
is formulated by a ``stochastization" \cite[Sect.~4.6]{Kubo_Toda_Hashitsume_1978}
of Navier-Stokes equation, and \emph{not} by coarse-graining Langevin equations for 
individual molecules in the fluid.  
The correct physical scenario to bear in mind is one of molecules 
squeezed together ``shoulder-to-shoulder" in a manner that allows only collective motion,
similar to that observed in a tightly packed crowd of people.  

\begin{figure}
\centering
\includegraphics[width=11cm]{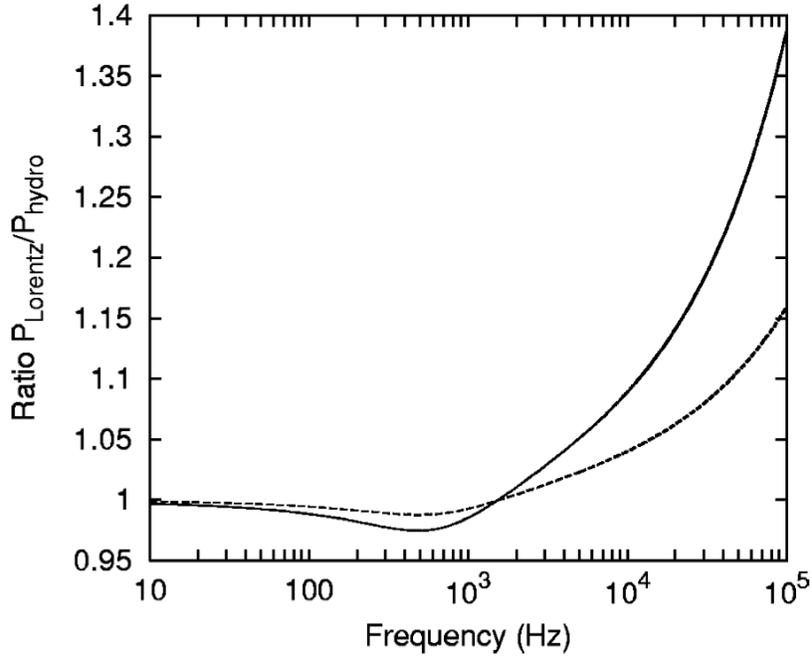}
\caption{The power spectrum of Brownian motion in an optical trap 
according to Einstein's theory, $P_{\rm Lorentz}$, 
divided by the hydrodynamically correct power spectrum for the same motion, $P_{\rm Hydro}$;
see \cite{RSI2004} for explicit expressions for the two spectra.
\emph{Fully drawn line:} Trap with Hooke's constant 3.8$\cdot$10$^{-2}$\,pN/nm 
for a micro-sphere with diameter 1\,$\mu$m.
\emph{Dashed line:} Hooke's constant 1.9$\cdot$10$^{-2}$ pN/nm for a micro-sphere with 
diameter 0.5\,$\mu$m.
At low time resolution, i.e., low frequency, the error vanishes.
Einstein made an excellent approximation when he chose Stokes' law 
for \textit{constant} velocity to characterize motion along a \textit{fractal} trajectory.}
\label{fig:1}       
\end{figure}

Equations~(\ref{eq:retardedfriction}--\ref{eq:hydronoise})
constitute the accepted hydrodynamically correct theory
for classical Brownian motion,
i.e., in an incompressible fluid.
It differs from Einstein's theory in a manner that matters
in practise with the precision that optical tweezers have 
achieved recently \cite{RSI2004,neuman_block_04};
see Fig.~\ref{fig:1}.


Power spectra of micro-spheres in optical traps can be 
measured with stochastic errors below 1\% \cite{RSI2004}.
So the differences in Fig.~\ref{fig:1} between Einstein's simple theory
and the hydrodynamically correct theory for Brownian motion in a fluid
can be exposed experimentally \cite{RSI2004,Petermann_et_al_2003,Lukic_et_al_2005}.  
The form of the thermal force in (\ref{eq:hydronoise}), on the other hand,
remains a theoretical result.
It is not a controversial result,
it is not questioned.
But because it is a small effect,
it has not yet been demonstrated experimentally.

\section{Power-Law Tails}

In the absence of external forces,
the position power spectrum of Brownian motion following from
(\ref{eq:retardedfriction}--\ref{eq:hydronoise})
is
\begin{equation} \label{eq:P}
P(f) \propto  \langle |\tilde{x}^2| \rangle  \propto
\frac{2\kT \Re \gstokes(f)}{|m(2\pi f)^2 +i2\pi f\gstokes(f)|^2}
\enspace.
\end{equation}
Here, the frequency-dependent numerator is the power spectrum of the
thermal force in (\ref{eq:hydro}),
while the denominator is given by the other terms in (\ref{eq:hydro}).
The frequency-dependent friction coefficient, $\gstokes(f)$,
appears both in numerator and denominator,
and both appearances contribute, with opposite signs,
to the $t^{-3/2}$ power-law tail in the velocity auto-correlation function.

By Wiener-Khintchine's theorem,
the velocity auto-correlation function is
\begin{equation}
\phi(t) = \langle \dot{x}(t) \dot{x}(0) \rangle \propto
\int_{-\infty}^{\infty} df e^{-i 2 \pi t f} (2\pi f)^2 P(f)
\enspace.
\end{equation}
At asymptotically large values of $t$, $\phi(t)$
is given by $P(f)$'s behavior at small  values of $f$,
\begin{equation}
(2\pi f)^2 P(f) = 2D( 1-(f/\fnu)^{1/2} + {\cal O}(f/\fnu)) \enspace.
\end{equation}
Hence
\begin{equation}
\phi(t) = \frac{D}{2\pi \fnu^{1/2}} t^{-3/2} + {\cal O}(t^{-5/2})
\mbox{~~~for~~~} t \rightarrow \infty   \enspace,
\end{equation}
quite different from the exponential decrease following from Einstein's
simple theory, but not conceptually different from it \cite{zwanzig_bixon_70,widom71,case71}.

Experimental evidence for this power-law tail remained sparse for years.
Dynamic light scattering offered promise of its observation,
but only Boon and Boullier \cite{Boon_Bouiller_1976,Bouiller_Boon_Deguent_1978}
reported an experimental result of the magnitude predicted theoretically,
with statistical errors about half the size of the signal.
Paul and Pursey used photon correlation dynamic laser light scattering
to measure the time dependence of the mean squared displacement
of polystyrene spheres with radius $R\sim 1.7\,\mu$m \cite{Paul_Pusey_1981}.
They found clear evidence for the expected $t^{-3/2}$-behavior
($t^{1/2}$ in the mean squared displacement),
but with an amplitude of only $74 \pm 3\%$ of that predicted theoretically.
They never found the reason why 26\%\ of the theoretically expected amplitude
was missing \cite{private_comm}.
Ohbayashi, Kohno, and Utiyama \cite{ohbayashi_etal_83}
also used photon correlation spectroscopy,
on a suspension of polystyrene spheres with radius 0.80\,$\mu$m,
and found agreement between the theoretical amplitude of the $t^{-3/2}$ tail
and their experimental results which has 9--10\%\ error bars.
Their results also agree with the predicted significant temperature dependence.
This convincing experiment thus supports the validity of the theory
(\ref{eq:hydro}).
This is the current experimental status of the power-law tail
of the velocity auto-correlation function of classical Brownian motion.

Or was, when this chapter was written.  
But before it went into print, Ref.~\cite{Lukic_et_al_2005} appeared.
Strangely, the velocity auto-correlation function is not given in \cite{Lukic_et_al_2005},
though its authors have measured what it takes to display its power-law tail.
Instead, they show the mean-squared-displacement of a diffusing micro-sphere.
That quantity is essentially the velocity auto-correlation function 
integrated twice, and
consequently contains the same power-law integrated twice.

The amplitude that was measured in all these experiments, 
albeit indirectly with photon correlation spectroscopy,
is  the first-order term in the expansion of $P(f)$ above, Eq.~(15), in 
powers of $(f/f_{\nu})^{1/2}$. This coefficient has two contributions: One from the 
denominator, from Stokes' frequency-dependent friction coefficient, and 
another from the numerator.
The latter is half-as-large as the former, 
and with opposite sign. 
It stems from the noise term's frequency dependence.

Instead of measuring a photon correlation function for laser light 
scattered off a suspension of micro-spheres,
developments in instrumentation \cite{neuman_block_04}
and data analysis \cite{RSI2004} for optical tweezers
have made it possible now to measure directly, with accuracy and precision,
on a \textit{single} micro-sphere \cite{Lukic_et_al_2005}.
Thus it just might be possible to observe directly the ''color" of the thermal noise,
the frequency dependence of the non-white power spectrum,
in a very challenging single-particle experiment with optical tweezers~\cite{NJP2005}.

 \section{In Situ Calibration of Optical Tweezers by Forced Nano-Scale Motion}
 
 There are many ways to calibrate an optical trap.
 Some ways are better than others if accuracy and precision is a concern.
 In that case, the best way is based on the motion's power spectrum \cite{RSI2004}.
 Two aspects must be calibrated: The spring constant of the Hookean force exerted by the
 trap on a trapped micro-sphere (bead), and, to this end, 
 the millivolt-to-nanometer calibration factor. 
 The latter tells us which nanometer-displacement of the bead in the trap corresponds 
 to a measured millivolt-change in output potential of a photo diode 
 in the position detection system used with the tweezers.
A common way to determine this calibration factor  
requires that one knows the radius of the bead, the temperature 
and dynamic viscosity of the fluid surrounding it,
and its distance to the nearby surface of the microscope cover slip,
if, as is usually the case, the experiment is done near this surface.
One can then calculate the bead's diffusion coefficient in m$^2/$s 
using Stokes' law (\ref{eq:stokeslaw}), Einstein's relation (\ref{eq:einsteinsrelation}), 
and Faxén's formula \cite{Faxen_1923,Happel_Brenner_1983}\cite[Sect.~XI]{RSI2004}.
By comparing the result with the same quantity measured experimentally in V$^2/$s,
the calibration factor is determined.

\begin{figure}
\centering
\includegraphics[width=11cm]{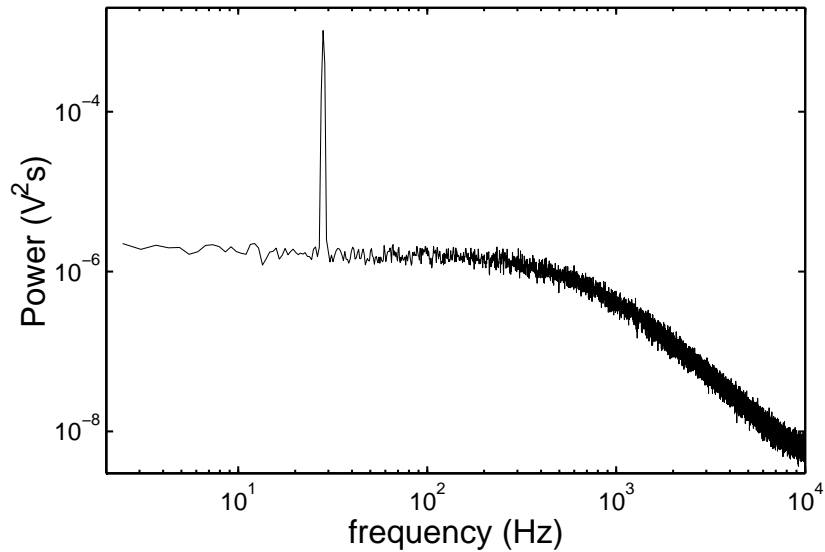}
\caption[]{Power spectrum of 1.54\,$\mu$m diameter silica bead 
held in laser trap with corner frequency $\fc=538$\,Hz.
The sample moves harmonically with amplitude $A=208$\,nm and frequency $\fstage=28$\,Hz.
The power spectrum shown is the average of 48 independent power spectra, 
sampled at frequency $\fsample=20$\,kHz.
The total sampling time was 79\,s, 
which is six times more than we normally would need to calibrate. 
It was chosen for the sake of illustration, 
to reduce the relative amplitude of the Brownian motion, 
i.e., the scatter in the spectrum \emph{away} from the spike at 28\,Hz.}
\label{fig:2}     
\end{figure}

A calibration of the photo diode that is much less dependent on a priori knowledge,
can be achieved  by moving the fluid cell with the bead harmonically 
relatively to the laboratory with the optical trap \cite{simon_erik_henrik2005}.
With a piezo-electric translation stage 
this can be done accurately with an amplitude of order 100\,nm
and frequency of order 30\,Hz.
In the laboratory system of reference,  the fluid
flows back and forth through the stationary trap with harmonically changing velocity.
This gives rise to an external force on the trapped bead in (\ref{eq:hydro}), 
a harmonically changing Stokes friction force,
\begin{equation} \label{eq:Fext}
\Fext(t)= \gamma_0 \vstage(t) = \gamma_0 2 \pi \fstage  A \cos(2 \pi \fstage(t-t_0))
\enspace,
\end{equation}
where $A$ and $\fstage$ are, respectively, 
the amplitude and frequency with which the stage is driven, 
and $t_0$ is its phase.
The amplitude $A$ can be chosen so small that the forced harmonic motion of the bead
in the trap  is masked by its Brownian 
motion, when observed in the time domain.
Nevertheless, when observed long enough, the forced harmonic motion stands out 
in the power spectrum of the total motion as a dominating  spike; see Fig.~\ref{fig:2}.
This spike is the dynamic equivalent of the scale bar plotted in micrographs:
The ``power"\ contained in it is known in m$^2$ because the bead's motion in nanometers
follows from its equation of motion and the known motion of the stage, measured in nanometers.
The bead's motion is \emph{measured} in Volts, however, by the photo-detection system,
and the Volt-to-meter calibration factor depends on the chosen signal amplification, 
laser intensity, etc.  So calibration is necessary.  
It is done by identifying the two values for the power in the spike:
The measured value in  V$^2$ with the known value in m$^2$ \cite{simon_erik_henrik2005}%
\footnote{A spike similar to the one shown here in Fig.~\ref{fig:2} 
is seen in \cite[Fig.~1b]{Svoboda_Schmidt_Schnapp_Block_1993}.
It was produced with a bead embedded in polyacrylamide, hence not moving thermally,
and not optically trapped.  
It was used to demonstrate the high sensitivity of the authors' position detection system.
It was also used for Volts-to-meters calibration of the detection system,
and gave 10\%\ agreement with the same calibration factor obtained from the power spectrum 
of Brownian motion.  
The optical properties of polyacrylamide differ from those of water, however,
so it is an open question how accurate that calibration method can be made.
Obviously, it is not an \emph{in situ} calibration method.}.
This method resembles an old method of calibration that 
moves the bead back and forth periodically with \emph{constant} speed,
but harmonic motion has a number of technical advantages.
One is that the precision of power spectral analysis demonstrated in \cite{RSI2004}
can be maintained, while adding the advantage of not having to know
the bead's radius, nor its distance to a nearby surface, nor the fluid's viscosity and temperature.
On the contrary, the combination of these parameters 
that occurs in the expression (\ref{eq:einsteinsrelation}) for the bead's diffusion coefficient, 
is determined experimentally from its Brownian motion, 
so, e.g., the bead's radius is measured 
to the extent the other parameter values are known. 
But also, this calibration method can be used \emph{in situ}, where an experiment is to be done,
by confining the bead's forced motion to this environment.
This is useful for measurements taking place near a surface, in a gel, or inside a cell.

\section{Biological Random Motion}

\begin{figure}
\centering
\includegraphics*[scale=0.5,angle=0]{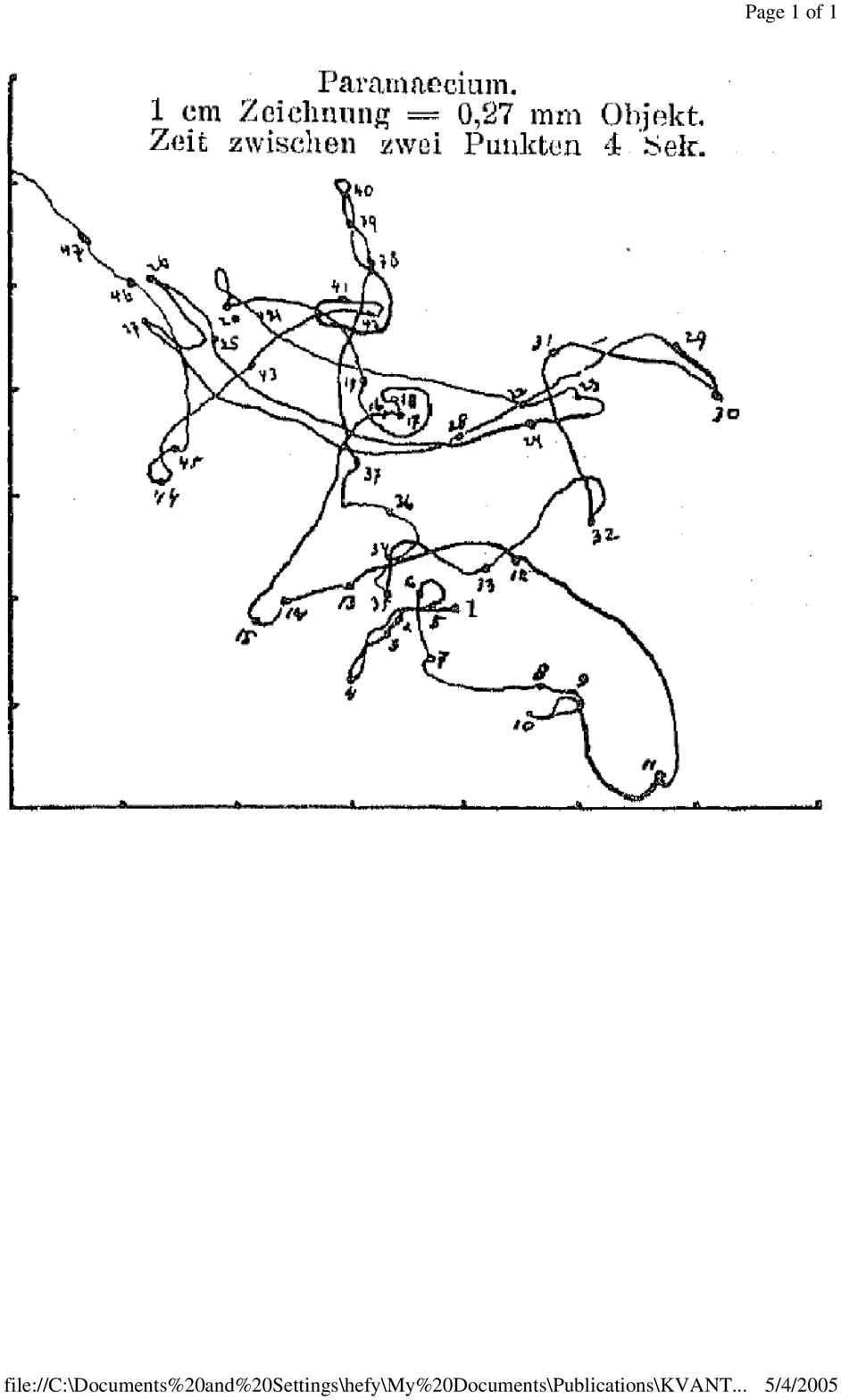} 
\caption[]{Example of Przibram's motility data, a trajectory of a protozoon, hand-drawn
with a mechanical tracking device operated in real time with a microscope.
A metronome was used to mark time on the trajectory every four seconds~\cite{przibram13}.}
\label{fig:3}     
\end{figure}

Robert Brown did not discover Brownian motion,
and he, a botanist, got his name associated with this physical phenomenon 
because he in 1827 carefully demonstrated what it is \emph{not}, a manifestation of life,
leaving the puzzle of its true origin for others to solve.
Brownian motion has been known for as long as the microscope,
and before the kinetic theory of heat
it was natural to assume that ``since it moves, it is alive."\ 
Brown killed that idea.
But after Einstein in 1905 had published his theory for Brownian motion,
Przibram in 1913 demonstrated  that this theory describes also the 
self-propelled random motion of protozoa \cite{przibram13}.
By tracking the trajectories $\vecx(t)$ of individual protozoa, see Fig.~\ref{fig:3},
Przibram demonstrated that the net displacement $\vecx(t)-\vecx(0)$ averages 
to zero, while its square satisfies the relationship known for Brownian motion,
\begin{equation}
\label{eq:rsquaredEinstein}
\langle \vecd(t)^2  \rangle= 2 \ndim D t
\enspace,
\end{equation}
where  $\ndim$ is the dimension of the space in which the motion takes place.

In Einstein's theory  $D$ is the diffusion coefficient,
and satisfies his famous relation (\ref{eq:einsteinsrelation}).
Przibram found a value  for $D$
which was much larger and much more sensitive to changes in temperature 
than Einstein's relation states.
He used this as proof that it was not just Brownian motion that he had observed.

If Przibram, a biologist, 
had used a better time resolution by marking out points in Fig.~\ref{fig:3}
more frequently than every four seconds, he might also have gotten ahead of 
the physicists in theoretical developments.
But he was drawing by hand, marking time to a metronome, so 
marking points closer to 1\,Hz must have been a challenge.

F\"urth, a physicist at the German university in Prague where Einstein had been a professor 
 for 16 months in 1911-12, also studied the motility of protozoa.
 First he repeated Przibram's results, apparently without knowing them \cite{furth17}.
Later he found that his data \cite{furth20} were \emph{not} 
described by  (\ref{eq:rsquaredEinstein}).
He consequently considered a random walker on a lattice,
and gave the walker directional persistence in the form of a bias towards 
stepping in the direction of the step taken previously.
By taking the continuum limit, he, 
independently of Ornstein \cite{ornstein18,uhlenbeck30}, demonstrated
that for random motion with persistence, (\ref{eq:rsquaredEinstein})
is replaced by 
\begin{equation}
\label{eq:dsquaredOU} \langle \vecd(t)^2  \rangle= 2 \ndim  D ( t - P (1
- e^{-t/P} )) \enspace,
\end{equation}
where  $P$ is called the \textit{persistence time},
and characterizes the time for which a given velocity is ``remembered"\ 
by the system \cite{furth20}.

Ornstein solved (\ref{eq:Langevin}),
since known as the Ornstein-Uhlenbeck (OU) process.
Its solution also gives (\ref{eq:dsquaredOU}), with $P \equiv m/\gamma$.
The physical meaning of the three terms in the OU-process
does not apply for cells:
Their velocities are measured in  \textit{micrometers} per \textit{hour}, 
so their inertial mass means absolutely nothing for their motion.
Friction with the surrounding medium also is irrelevant---the cells are 
firmly attached to the substrate they move on---and it is not thermal forces that accelerate the cells.
But as a mathematical model the OU-process is the simplest possible of its kind,
like the harmonic oscillator, the Hydrogen atom, and the Ising model.
It also agrees with the earliest data.
Consequently, the OU-process became the standard model for motility.
We can write it as 
\begin{equation} \label{eq:OU}
P \frac{d\vecv}{dt} = - \vecv + (2D)^{1/2} \veceta \enspace,
\end{equation}
where each component of $\veceta$ is a white noise normalized as in (\ref{eq:whitenoise}) and uncorrelated with the other components.
\begin{equation}\label{noisedef}
 \langle \veceta(t)\rangle=\vecnull ~~;~~~
\langle \eta_j(t') \eta_k(t'')\rangle= \delta_{j,k}\delta(t'-t'') \enspace.
\end{equation}
Here $\delta(t)$ and $\delta_{j,k}$ are, respectively Dirac's and Kronecker's $\delta$-functions, and $\veceta(t)$ is assumed uncorrelated with $\vecv(t')$ for $t\ge t'$.
F\"urth's formula (\ref{eq:dsquaredOU}) is a consequence of equations~(\ref{eq:OU}) and (\ref{noisedef}), but follows also from other, similar theories.
It was often the only aspect of the theory that was compared with experimental data,
and with good reason, considering the limited quality of data.

Gail and Boone \cite{gail70} seem to have been the first to model cell motility
with (\ref{eq:dsquaredOU}).
They did a time study of  fibroblasts from mice by measuring the cells' positions every 2.5\,hrs.
Equation~(\ref{eq:dsquaredOU}) fitted their results fairly well.
Since then, cell motility data have routinely been fitted with  (\ref{eq:dsquaredOU}).
Its agreement with data can be impressive, and is usually satisfactory%
---sometimes helped by the size of experimental error bars and few points at times $t$ that are comparable to  $P$.
Data with these properties cannot distinguish (\ref{eq:dsquaredOU}) 
from other functions that quickly approach $2 \ndim D ( t - P)$.

Equation~(\ref{eq:dsquaredOU}) is essentially a double integral of the velocity
auto-correlation function $\phi(t)$ of the OU-process, where
\begin{equation}\label{funeral}
 \phi(t)= \langle \vecv(0) \cdot \vecv(t)\rangle=\frac{\ndim D}{P}e^{-|t|/P}
  \enspace.
\end{equation}
Experimental results for the velocity auto-correlation function 
are better suited for showing whether the OU-process is a reasonable model for given data.
But experimental results for velocities are calculated as finite differences 
from  time-lapse recordings of positions. 
If the time-lapse is short, precision is low on differences, hence on computed velocities.
Yet, if the time-lapse is longer, the time resolution of the motion is poor.
The solution is somewhere in between, compensating for lost precision with good statistics.
Good statistics was not really achievable till computer-aided object-tracking became possible.

\begin{figure}
\centering
\includegraphics[width=5.8cm]{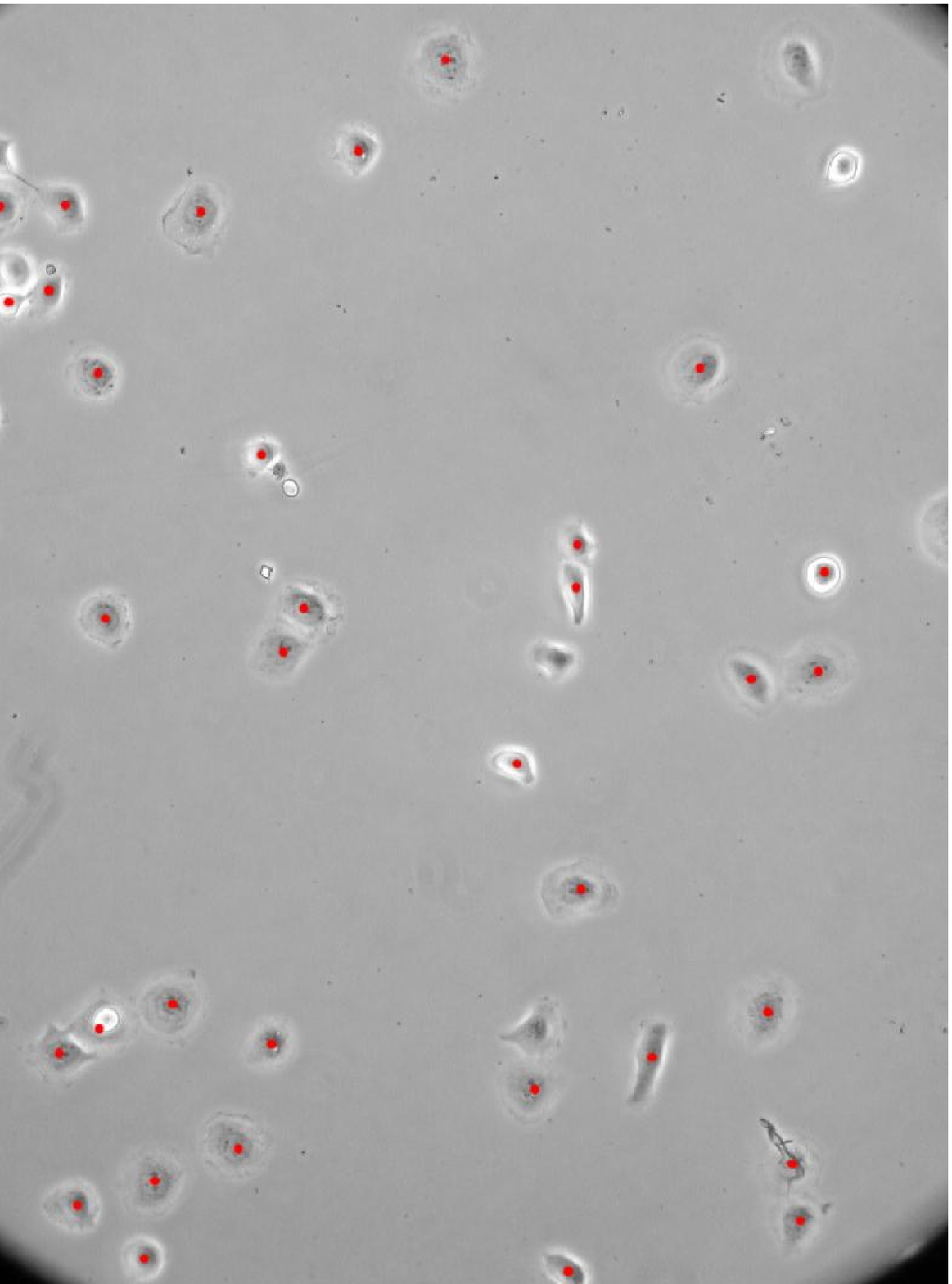}
\includegraphics[width=5.4cm]{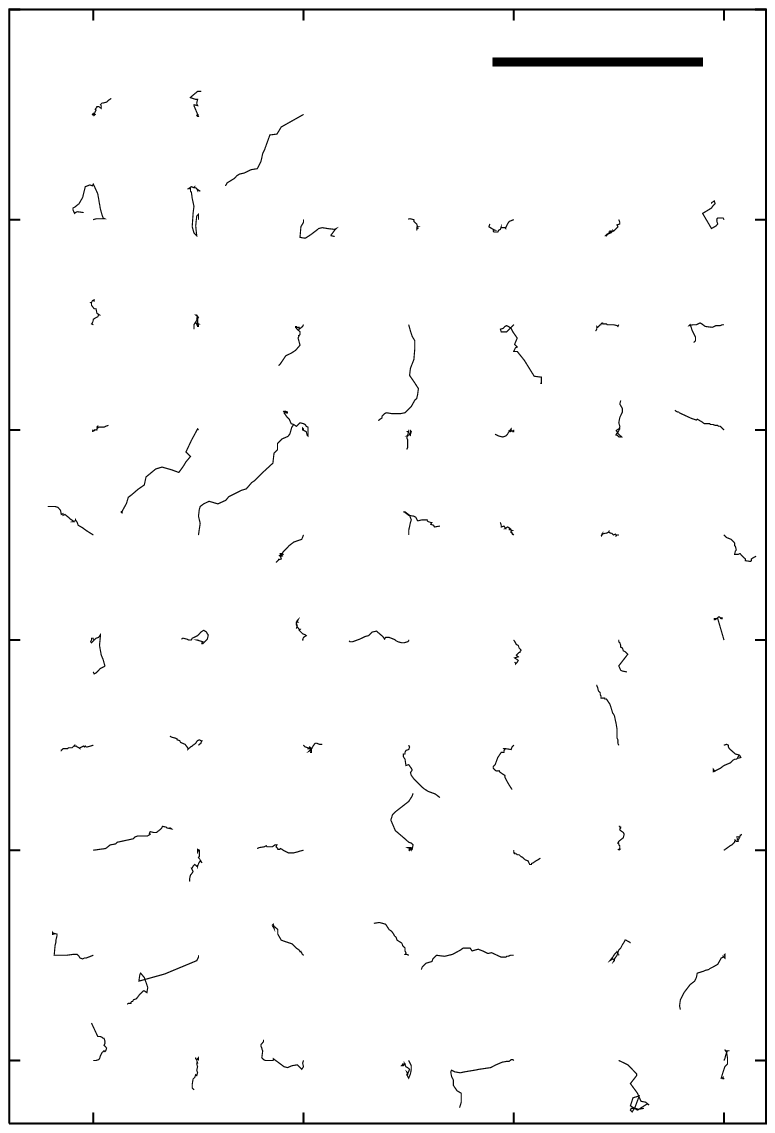}
\caption[]{\label{fig:4}
Isolated human dermal keratinocytes are  motile by nature.
If not surrounded by other cells, they react as if in a wound:
They search for other cells of the same kind with which they can connect to form skin.
Trajectories are formed from 15\,min time-lapse photography.
Trajectories as those shown here in the right panel make up the raw data that 
are analyzed statistically to find a suitable stochastic model 
for the motility of these keratinocytes.
The black bar is 0.2\,mm long. }
\end{figure}

\section{Enter Computers}

We recently wanted to characterize the compatibility of human cells with various surfaces
by describing the cells' motility on the various surfaces \cite{BJ2005}.
Computer-aided cell tracking---see Fig.~\ref{fig:4}---quickly gave us so much data 
that we found ourselves in a new situation with regards to modelling:
We were not limited to showing whether or not there is agreement between data 
and a few consequences of a given model.
We could investigate the model itself experimentally,
measure each term in its defining equation, check that their assumed properties 
are satisfied,
and whether together they satisfy the equation of motion.

Furthermore, before we checked the equation of motion, we could check
whether the data are consistent with various assumptions of symmetry and invariance 
on which the equation of motion is based.
We found that the cells behaved in a manner consistent with 
the assumptions that their surroundings are isotropic, homogenous, and constant in time.
This allowed us to average data over all directions, places, and times.
This in turn improved the statistics of our investigation of the equation of motion \cite{BJ2005}.


\section{Tailor-Made Theory Replaces ``One Theory Fits All"}

The theory in (\ref{eq:OU}) states that for a given velocity $\vecv$
the acceleration is a stochastic variable with expectation value  proportional to $\vecv$,
\begin{equation}
 \langle \frac{d\vecv}{dt} \rangle_{\vecv} = -  \vecv/P  \enspace.
\end{equation}
Figure~\ref{fig:5}AB shows that this is also the case for experimental data.

\begin{figure}
\centering
\includegraphics[width=5.45cm]{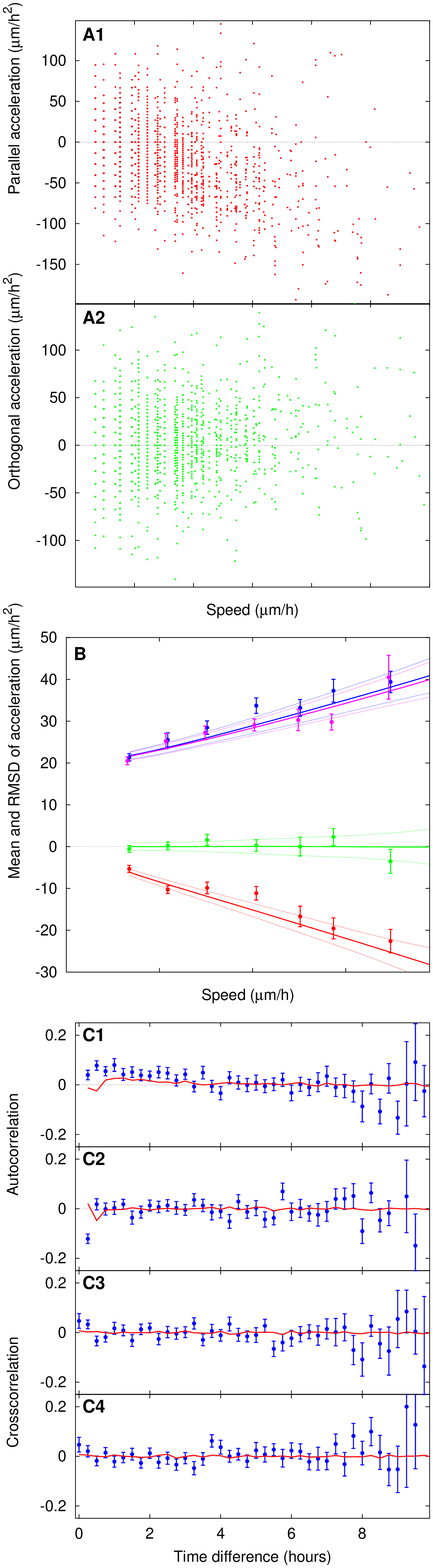}
\includegraphics[width=5.55cm]{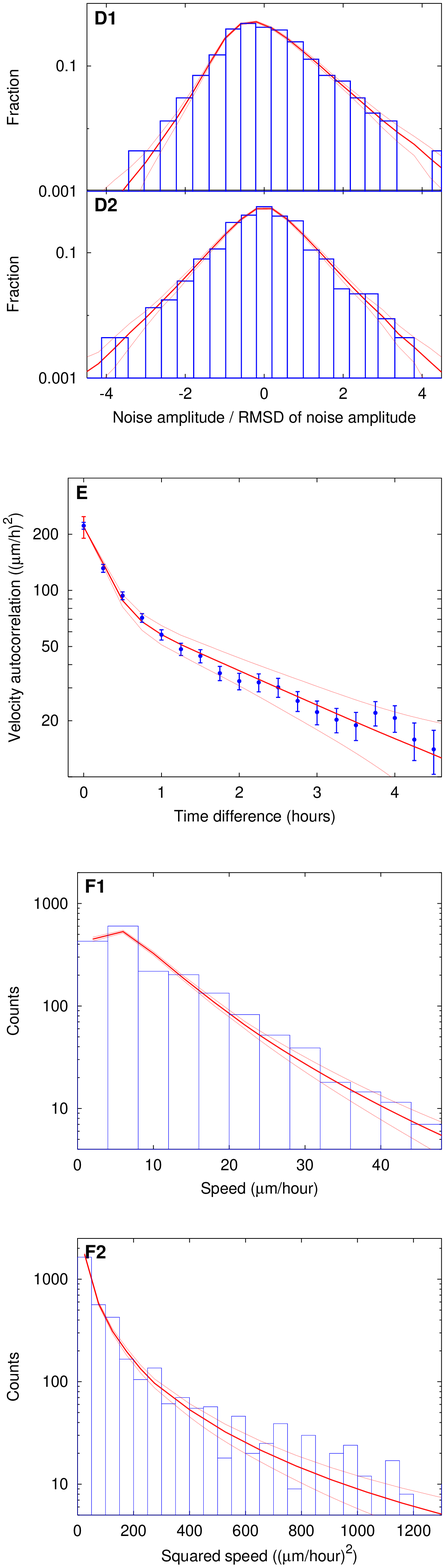}
\end{figure}

\begin{figure}
\caption{\label{fig:5} 
Statistics accumulated from trajectories like those shown in Fig.~\ref{fig:4}.
\textbf{A:}
The two components of the acceleration, as functions of speed.
Panels~A1 and A2 show the acceleration parallel with, respectively orthogonal to, the velocity.
These scatter plots show that the two functions contain random parts,
like the acceleration in (\ref{eq:OU}).
\textbf{B:} Data points with error bars: Mean and standard deviation as function of speed
for data shown in Panel~A\@.
Curves show the same quantities,  plus/minus one  standard deviation, 
calculated from the theory given in (\ref{eq:HaCaT}).
\textbf{C:} Correlation functions for scatter shown in Panel~A\@.
 Panels C1 and C2 show the auto-correlations of the two components,
C3 and C4 show the cross-correlation between the two, for both signs of the time difference.
The many values shown are almost all indistinguishable from zero.
This suggests that the scatter in data can be modelled with uncorrelated noise,
as in (\ref{noisedef}).
This is an experimental result for the theory we seek.
The curves shown are not fits to the data shown, 
but results of (\ref{eq:HaCaT}) after it has been fitted to data in Panels~B, E, and F\@.
\textbf{D:}~Histograms of scatters shown in Panel~A, 
measured relatively to the means shown in Panel~B, and in units of
the standard deviations shown in Panel~B\@.
The curves shown are not fits to the histograms shown, 
but results of (\ref{eq:HaCaT}) after it has been fitted to data in Panels~B, E, and F\@.
\textbf{E:}~Velocity auto-correlation function, 
calculated from trajectories like those shown in Fig.~\ref{fig:4}.
It is not a simple exponential as in (\ref{funeral}).
But a sum of two exponentials fit data perfectly.
So we assume that the theory we seek has a velocity auto-correlation function
that is a sum of two exponentials.
The curves through the data points are that correlation function,  plus/minus one
 standard deviation, computed with the theory in (\ref{eq:HaCaT}), 
 after it has been fitted to the data shown here, and simultaneously to the data 
 in Panels~B and F\@.
\textbf{F:}  Histograms of speeds and  (speed)$^2$
read off trajectories like those in Fig.~\ref{fig:4}.
The curves shown are the same speed distributions calculated 
from the theory in (\ref{eq:HaCaT}),
after it has been fitted to the data.}
\end{figure}

The theory in (\ref{eq:OU}) states also that
\begin{equation}
 \frac{d\vecv}{dt}-\langle \frac{d\vecv}{dt} \rangle_{\vecv} 
 =  \frac{d\vecv}{dt} + \vecv/P  = (2D)^{1/2} \veceta / P \enspace,
\end{equation}
i.e., that this quantity in the OU-process is a white noise 
with the same speed-independent amplitude in both directions: 
parallel and orthogonal to the velocity.

Figure~\ref{fig:5}B shows that experimentally the amplitude of the two components of this noise
are indeed indistinguishable in the two directions, 
but the two amplitudes are clearly  \emph{not} independent of the speed!
Here we see the experimental data reject the OU-process as model.
The distribution of experimentally measured values of the noise 
also reject the OU-process as model.
Figure~\ref{fig:5}D shows clearly that it is \emph{not} Gaussian, as it is in the OU-process.
Apart from that, Fig.~\ref{fig:5}C shows 
that the noise is uncorrelated, like in (\ref{noisedef}),
on the time scale where we have measured it.
This result radically simplifies the mathematical task of constructing an alternative to
the OU-process on the basis of experimentally determined properties of these cells' motility pattern.

The velocity auto-correlation function of the OU-process is a simple exponential,
(\ref{funeral}).
Figure~\ref{fig:5}E shows the experimentally measured velocity auto-correlation function.
It is fitted perfectly by the  \textit{sum} of two exponentials,
so again the experimental data reject the OU-process as model.

The data shown in Fig.~\ref{fig:5} are so rich in information that with a few assumptions
favored by Occam's Razor one can deduce \emph{from the data} 
which theory it takes to describe the data, 
and this theory is unambiguously defined by the data~\cite{BJ2005}.
Results from this theory are shown as the fully drawn curves passing through the data points 
in Fig.~\ref{fig:5}.
It is given by the stochastic integro-differential equation
 \begin{eqnarray}
\lefteqn{\frac{d\vecv}{dt}(t) = -\beta \, \vecv(t)} &&       \label{eq:HaCaT}\\
&& +\alpha^2 \int_{-\infty}^t dt' e^{-\gamma(t-t')}\vecv(t')
+\sigma(v(t)) \, \veceta(t)
\enspace, \nonumber
\end{eqnarray}
where
\begin{equation}  \label{eq:sigmavHaCaT}
\sigma(v)=\sigma_0 + \sigma_1 v  \enspace.
\end{equation}
The integral over past velocities in (\ref{eq:HaCaT}) 
is called a memory-kernel by mathematicians. 
It shows that these cells have memory.
This is no surprise: The polarity of the cytoskeleton of a moving cell
is a manifest memory of direction, and while its instantaneous velocity depends on 
the activity of transient pseudopodia, 
the fact that pseudopodia are active depends on states of the cell
that last longer than the individual pseudopod, one would expect.

Note the similarity between (\ref{eq:HaCaT}) 
and the hydrodynamically correct theory for Brownian motion, 
(\ref{eq:retardedfriction},\ref{eq:hydro}).   
Though both are more complex than Einstein's theory, applied by Przibram,
and the OU-process, applied by Fürth, they still have much in common.
This is so because they both are \emph{linear} 
and both respect \emph{causality} and the same \emph{space-time symmetries}.

Linearity is simplicity, so wherever in modelling it is sufficient, one avoids going beyond it.
This is why (\ref{eq:retardedfriction},\ref{eq:hydro}) and (\ref{eq:HaCaT}) are both linear.

The \emph{Principle of Causality} states that the future does not affect the present, 
including present rates of change of state variables.  
Only the past can do this.
This principle is respected throughout physics, 
and we have of course built it into our motility models as well.
This is why the rate of change of the velocity 
given in (\ref{eq:retardedfriction},\ref{eq:hydro}), respectively (\ref{eq:HaCaT}), 
depends only on past and present velocities.
The integral kernels occurring in both equations are \emph{memory} kernels
in order to respect this principle.

In a homogenous, isotropic environment that is constant in time, 
there is no absolute position, direction, nor time.   
A theory for a dynamical system in such an environment
consequently cannot depend on the position variable $\vecx$, 
nor can it depend explicitly on the time variable $t$, 
nor on explicit directions in space.
The theory must be \emph{translation invariant} in space, time, and with respect to direction.
The last invariance is called \emph{covariance under rotations}, because 
a theory for a vector variable like the velocity is not invariant under rotations of
the coordinate system, it is covariant, i.e., transforms like the vector it describes.
Because these space-time symmetries are shared by hydrodynamics and our cells, 
neither (\ref{eq:retardedfriction},\ref{eq:hydro}) nor (\ref{eq:HaCaT})
depends on $\vecx$, nor explicitly on $t$, and both models transform like a vector under 
rotations.

We conclude that with the rich data that one now can record and process,
one should not be satisfied with the simplest possible model for persistent random motion, 
the OU-process.  ``One size fits all"\ is no longer true, if it ever was.
Motility models can be made to measure.
Here we have only presented the first phenomenological steps of that process:
How to plot and read motility data in a manner 
that reveals mathematical properties of the theory sought.
That done, it is another task to construct a model with the properties demanded.
If that can be done, it is yet another task to decide whether the theory is unique or not.
Two examples of such theories and their derivation are given in \cite{BJ2005}.


\end{document}